\documentclass[%
reprint,
superscriptaddress,
showpacs,
preprintnumbers,
bibnotes,
amsmath,amssymb,
aps,
]{revtex4-1}

\usepackage{graphicx}
\usepackage{dcolumn}
\usepackage{bm}
\usepackage{chemformula}
\usepackage{hyperref}

\usepackage{color}

\usepackage{tabularx}
\usepackage{amsmath}
\usepackage{float}

\newcommand{\parallelsum}{\mathbin{\!/\mkern-5mu/\!}}

\begin{document}

\title{Paramagnons and high-temperature superconductivity in mercury-based cuprates}

\author{Lichen~Wang}
\thanks{These authors contributed equally to this study.}
\affiliation{International Center for Quantum Materials, School of Physics, Peking University, Beijing 100871, China}
\affiliation{Max Planck Institute for Solid State Research, Stuttgart 70569, Germany}
\author{Guanhong~He}
\thanks{These authors contributed equally to this study.}
\affiliation{International Center for Quantum Materials, School of Physics, Peking University, Beijing 100871, China}
\author{Zichen~Yang}
\affiliation{Max Planck Institute for Solid State Research, Stuttgart 70569, Germany}
\author{Mirian~Garcia-Fernandez}
\affiliation{Diamond Light Source, Harwell Campus, Didcot OX11 0DE, United Kingdom}
\author{Abhishek~Nag}
\affiliation{Diamond Light Source, Harwell Campus, Didcot OX11 0DE, United Kingdom}
\author{Ke-Jin~Zhou}
\affiliation{Diamond Light Source, Harwell Campus, Didcot OX11 0DE, United Kingdom}
\author{Matteo~Minola}
\affiliation{Max Planck Institute for Solid State Research, Stuttgart 70569, Germany}
\author{Matthieu~Le~Tacon}
\affiliation{Institute for Quantum Materials and Technologies, Karlsruhe Institute of Technology, Karlsruhe 76133, Germany}
\author{Bernhard~Keimer}
\affiliation{Max Planck Institute for Solid State Research, Stuttgart 70569, Germany}
\author{Yingying~Peng}
\email[]{yingying.peng@pku.edu.cn}
\affiliation{International Center for Quantum Materials, School of Physics, Peking University, Beijing 100871, China}
\affiliation{Collaborative Innovation Center of Quantum Matter, Beijing 100871, China}
\author{Yuan~Li}
\email[]{yuan.li@pku.edu.cn}
\affiliation{International Center for Quantum Materials, School of Physics, Peking University, Beijing 100871, China}
\affiliation{Collaborative Innovation Center of Quantum Matter, Beijing 100871, China}

\date{\today}

\begin{abstract}
We present a comparative study of magnetic excitations in the first two Ruddlesden-Popper members of the Hg-family of high-temperature superconducting cuprates, which are chemically nearly identical and have the highest critical temperature ($T_\mathrm{c}$) among all cuprate families. Our inelastic photon scattering experiments reveal that the two compounds’ paramagnon spectra are nearly identical apart from an energy scale factor of $\sim130\%$ that matches the ratio of $T_\mathrm{c}$'s, as expected in magnetic Cooper pairing theories. By relating our observations to other cuprates, we infer that the strength of magnetic interactions determines how high $T_\mathrm{c}$ can reach. Our finding can be viewed as a magnetic analogue of the isotope effect, thus firmly supporting models of magnetically mediated high-temperature superconductivity.

\end{abstract}

\maketitle

Identifying the driving force for Cooper pairing in cuprate high-temperature superconductors is an outstanding quest in quantum materials research \cite{Keimer2015}. Magnetic interactions are widely considered to play a key role \cite{Lee2006,Scalapino2012}, yet a proof of this demands establishing a correspondence between the strength of Cooper pairing and the energy of magnetic excitations, known as paramagnons \cite{LeTacon2011} in cuprates doped away from their parent antiferromagnetic state. Indeed, such a correspondence would parallel the isotope effect \cite{Maxwell1950,Reynolds1950}, which identified phonons as mediators of Cooper pairing in conventional low-temperature superconductors. There, because phonon frequencies (or energies) set a limit on the pairing strength, $T_\mathrm{c}$ varies as the inverse square root of the atomic mass $m$ of the isotope, given that the isotopic replacement affects the phonon frequencies as $1/\sqrt{m}$ but leaves all other properties intact to an excellent approximation.

To test the idea that superconductivity in the high-$T_\mathrm{c}$ cuprates is magnetically mediated \cite{Lee2006,Scalapino2012}, a similar relation between $T_\mathrm{c}$ and magnetic energies would provide strong empirical evidence. As antiferromagnetic spin fluctuations manifest themselves as paramagnons in superconducting cuprates \cite{LeTacon2011}, the paramagnon energies are expected to scale with $T_\mathrm{c}$ in models where Cooper pairing is mediated by the spin fluctuations \cite{LeTacon2011,Dahm2009}. For a similar reason, the discovery of the so-called spin resonant modes, first in the cuprates \cite{RossatMignod1991,Eschrig2006} and then in heavy-fermion \cite{Sato2001,Stock2008} and iron-based \cite{Christianson2008} superconductors, has stimulated intense research interest -- the energy of the resonant mode is often found to be proportional to $T_\mathrm{c}$ \cite{Eschrig2006,Yu2009}. However, the resonant mode is present only in the superconducting state and its energy falls below the superconducting energy gap $2\Delta_\mathrm{SC}$ \cite{Yu2009}, thus its formation can neither drive nor set an energy limit on the Cooper pairing, and is likely a consequence of pairing \cite{Scalapino2012}.

The pursuit for an energy correspondence between pairing and the paramagnons has proved challenging. Superconductivity is known to exist in a dome-like region of the generic $p$-$T$ phase diagram \cite{Keimer2015,Lee2006,Scalapino2012} of the cuprates, where $T$ is temperature, and $p$ is the carrier concentration introduced into the charge-transfer insulating, antiferromagnetic parent compounds. The decrease of $T_\mathrm{c}$ away from optimal doping is believed to be due to reduced superfluid density \cite{Emery1995,Bozovic2016}, whereas the paramagnon energies are nearly independent of $p$ \cite{LeTacon2011,Peng2018,LeTacon2013,Dean2013,Jia2014}. Thus, the hope for finding the correspondence falls upon using a tuning knob other than doping, such as chemical \cite{Ofer2006} and applied pressures \cite{Mallett2013}. Even though a positive correlation between $T_\mathrm{c}$ and parent compounds’ antiferromagnetic ordering temperature has been found in some cases \cite{Ofer2006}, spectroscopic determination of the associated energies suggests a much weaker correspondence \cite{Ellis2015}, sometimes with controversial results \cite{Mallett2013,Dean2014,Peng2015}. This is because major modifications of the magnetic exchange interactions ($J$) are difficult to achieve \cite{Keren2019}, and they go along with structural modifications whose consequences are difficult to assess. In particular, most of the materials previously studied in this regard have relatively low $T_\mathrm{c}$, making it unclear whether the variation in $T_\mathrm{c}$ is linked to the magnetic energy or some detrimental effects on $T_\mathrm{c}$, including chemical disorder \cite{Eisaki2004} and competing states such as charge order \cite{Keimer2015}, which may vary at the same time as the tuning takes place. These limitations have prevented the previous experimental indications from being widely recognized as establishing a magnetic-pairing mechanism.

\begin{figure*}
	\centering{\includegraphics[clip,width=16cm]{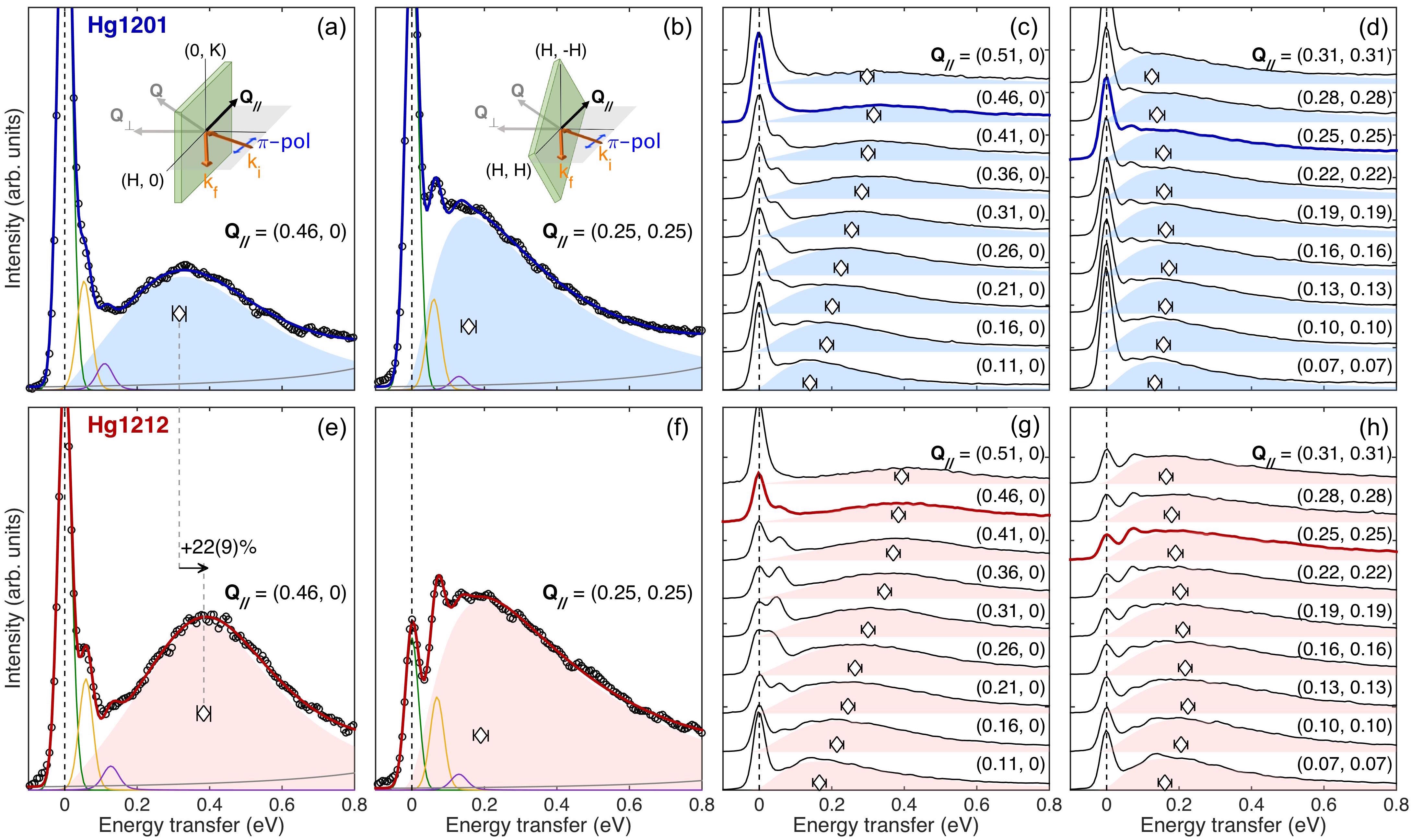}}
	\caption{RIXS spectra measured at $T = 13$ K with $\pi$-polarized incident x-rays. (a)-(b), Representative data measured on Hg1201 with in-plane momentum transfers $\mathbf{Q}_{\parallelsum} = (0.46, 0)$ and $(0.25, 0.25)$, respectively, in reciprocal lattice units (r.l.u.). The data are normalized to the intensity of $dd$ excitations at higher energies (Fig. S2 in \cite{SM}), and then fitted to a sum of an elastic peak (green), a single-phonon peak (yellow), a two-phonon peak (magenta), a paramagnon peak (shaded area, diamond symbol indicates the peak maximum), and a weakly energy-dependent background (grey). Details of the fitting are presented in \cite{SM}, and the results are shown in Figs. S3-6 and Table S1 in \cite{SM}.
Insets illustrate the scattering geometry, where the detector is placed at a fixed $2\theta$ angle of 154$^\circ$ from the incident beam. The desired $\mathbf{Q}_{\parallelsum}$ is reached by rotating the sample around the vertical axis. (c)-(d), Spectra at a series of $\mathbf{Q}_{\parallelsum}$ along high-symmetry directions, vertically offset for clarity. The fitted paramagnon peaks are displayed by shaded areas along with the data. (e)-(h), Same as (a)-(d), but for Hg1212. Error bars represent uncertainty in the estimate of the energies of the paramagnon signal’s intensity maxima (1 s.d.), see \cite{SM} for details.}
	\label{fig1}
\end{figure*}

To attain a definitive answer on this front, it is desired to study materials with very high yet sufficiently different $T_\mathrm{c}$, so that $T_\mathrm{c}$ is not strongly reduced by material-specific details and its variations can be unambiguously compared to that of the magnetic energy. The first two Ruddlesden-Popper (RP) members of the Hg-family of cuprates, HgBa$_2$CuO$_{4+\delta}$ (Hg1201) and HgBa$_2$CaCu$_2$O$_{6+\delta}$ (Hg1212), are ideal for such a test of the magnetic ``isotope effect'' on high-$T_\mathrm{c}$ superconductivity. Hg1201 and Hg1212 are structurally and chemically nearly identical, have the highest $T_\mathrm{c}$ (97 K and 127 K at optimal doping, respectively) among all cuprate families \cite{Jorgensen2000,Eisaki2004,Barisic2008}, and all CuO$_2$ layers are identical by symmetry. For later RP members with three or more consecutive CuO$_2$ layers, the charge imbalance between the layers \cite{Chakravarty2004} complicates the analysis. Previously, the difference in $T_\mathrm{c}$ between Hg1201 and Hg1212 has been attributed to quantum tunnelling of Cooper pairs between the adjacent CuO$_2$ layers \cite{Chakravarty2004}. But as we will show, this understanding has missed a distinct variation in the magnetic energy within the individual CuO$_2$ layers, a variation that is strong enough to mostly account for the change in the pairing strength.

We have performed inelastic photon scattering experiments on two nearly equally underdoped high-quality single crystals of Hg1201 and Hg1212 (Fig. S1 in \cite{SM}), with $T_\mathrm{c}$ of 80 K ($p\sim0.11$) and 107 K ($p\sim0.12$), respectively. Our primary measurement technique is resonant inelastic x-ray scattering (RIXS). By using incident x-ray photons tuned to the energy of the $L_3$ absorption edge of Cu$^{2+}$, we are able to directly probe single spin-flip excitations on the Cu$^{2+}$ square lattice \cite{Ament2011}. The excitations in doped cuprates are called paramagnons \cite{LeTacon2011}, after (single) magnon excitations in the magnetically ordered parent compounds. Over the past two decades, the remarkable development in soft-x-ray instrumentation \cite{Ament2011} has made RIXS an effective method for determining the energy- and momentum-dependent paramagnon spectrum in the cuprates \cite{LeTacon2011,Peng2018,LeTacon2013,Dean2013,Jia2014,Braicovich2010}. The dominance of these collective excitations over incoherent particle-hole excitations in the RIXS spectra has been demonstrated up to at least optimal doping $p\sim0.16$ in various cuprates \cite{Minola2015,Minola2017}.

\begin{figure*}
	\centering{\includegraphics[clip,width=16cm]{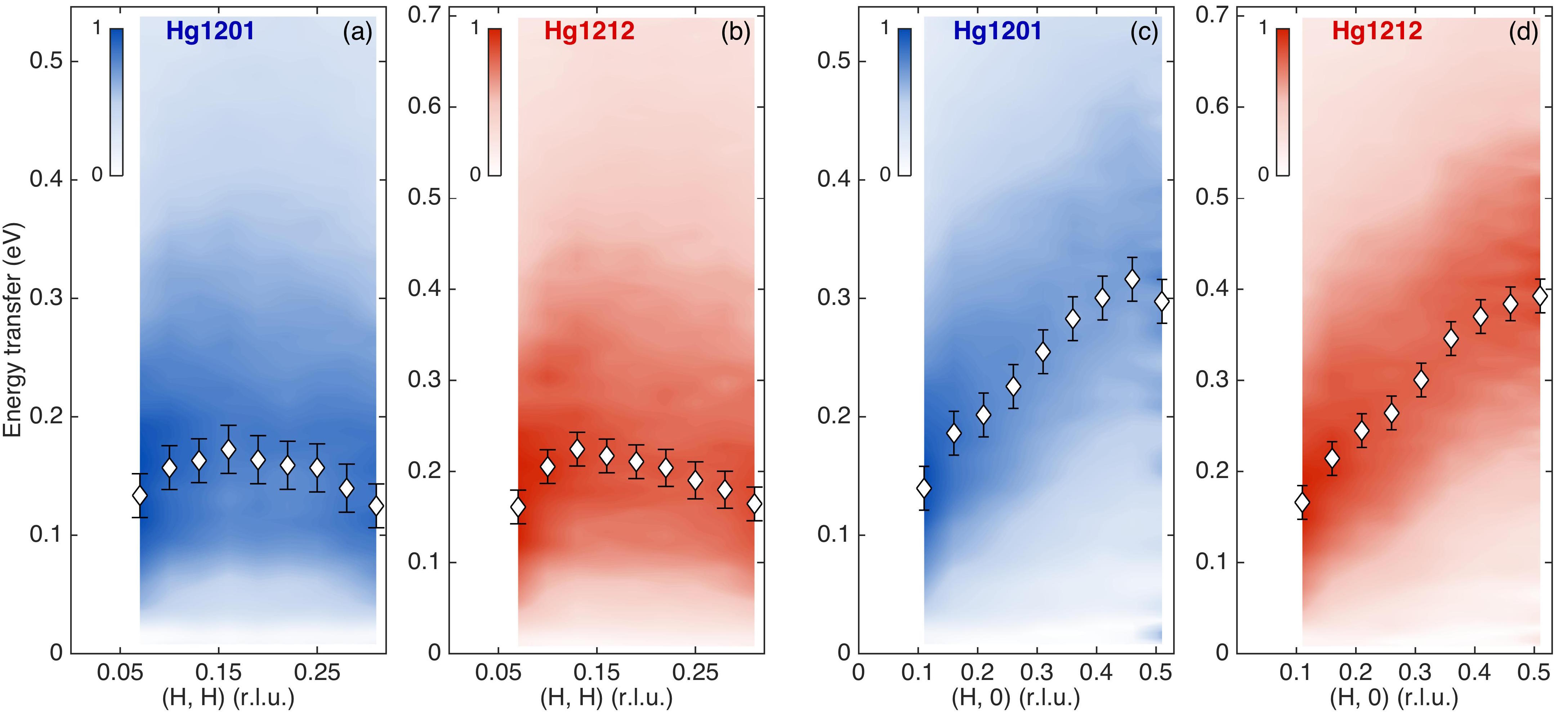}}
	\caption{Comparison of paramagnons in Hg1201 and Hg1212. (a)-(b), False-colour representation of energy- and momentum-dependent RIXS intensities arising from paramagnons in Hg1201 and Hg1212, respectively, along the $(H, H)$ momentum direction. The signal is extracted from the data displayed in Fig. 1d and h, after subtracting the non-magnetic contributions and normalizing the energy-integrated spectral weight among different $\mathbf{Q}_{\parallelsum}$. Diamond symbols indicate energy positions of intensity maxima (according to the DHO fits in Fig. 1) at the measured $\mathbf{Q}_{\parallelsum}$, with values and error bars identical to those in Fig. 1. (c)-(d), Same as (a)-(b), but for the $(H, 0)$ momentum direction. For visual comparison between the two systems, the vertical energy scales of (b) and (d) are set to be $130\%$ those of (a) and (c).}
	\label{fig2}
\end{figure*}

Figure 1 displays our representative RIXS spectra obtained with $\pi$-polarized incident photons. The in-plane momentum transfer $\mathbf{Q}_{\parallelsum}$ is sampled along the high-symmetry lines $(H, 0)$ and $(H, H)$ of the magnetic Brillouin zone. In the displayed energy range, the RIXS intensity consists of five components: elastic, single- and two-phonon scattering, a weakly energy-dependent background, and magnetic scattering mainly from paramagnons \cite{LeTacon2011,Peng2018}. A satisfactory account for the measured intensity is obtained with such a model (Fig. 1a, b, e, f, see \cite{SM} for details), where the paramagnon signal is described by a damped harmonic oscillator (DHO) peak, visualized by shaded areas in Fig. 1. This peak’s energy clearly disperses along the $(H, 0)$ direction (Fig. 1c and g), and a comparison between the two systems close to the Brillouin zone boundary at $H = 0.46$ indicates a distinct energy increase from Hg1201 (Fig. 1a) to Hg1212 (Fig. 1e), which amounts to $22\pm9\%$ of the energy in Hg1201 as determined from the fit maximum of the DHO peak. A similar energy increase is seen from the fits also in the $(H, H)$ direction (Fig. 1b and f), but the comparison is less obvious from the raw data due to the overdamped nature of the excitations.

The above observation calls for a systematic comparison between the two systems concerning the entire set of the paramagnon spectra, which we present in detail in Figs. S3-6 and Tables S1-2 in \cite{SM}. Our conclusion is that, from Hg1201 and Hg1212, the propagation energies of the paramagnons \cite{LeTacon2011,Peng2018} increase globally by approximately $30\%$. To visualize the increase, we plot in Fig. 2 the energy- and momentum-dependent RIXS intensities, after removing the non-magnetic contributions and normalizing the integrated spectral weight at each $\mathbf{Q}_{\parallelsum}$. In the panels for Hg1212 (Fig. 2b and d), the vertical energy scales are purposely set to be $130\%$ those of the Hg1201 panels (Fig. 2a and c). It is clear that the results are very similar apart from the energy-scale difference. A pursuit for the best visual similarity would instead suggest a rescaling ratio of about $122\%$ (Fig. S7 in \cite{SM}), close to the comparison in Fig. 1a and e. Here we consider the ratio associated with the propagation energy physically more meaningful, since visual comparison may be biased by the effect of damping in the DHO profiles, and because the propagation energy is known to change little with doping \cite{LeTacon2011,Peng2018,LeTacon2013,Dean2013,Jia2014}, making it suitable for representing the strength of magnetic interactions in a given system.

Importantly, the 22-30$\%$ increase in the paramagnon energies from Hg1201 to Hg1212 is enough to account for most, if not all, of the difference in $T_\mathrm{c}$ between the two systems both at optimal doping (from 97 K to 127 K) and in our actual samples (from 80 K to 107 K). The increase must originate in the strength of magnetic interactions within the CuO$_2$ layers, since magnetic coupling between adjacent CuO$_2$ layers in a bilayer cuprate is only about 10 meV, which has little effect on the (para)magnon energies far away from the zone center \cite{Reznik1996}. Indeed, if one takes $T_\mathrm{c}$ as the measure for the pairing strength, our result readily constitutes strong evidence for a large magnetic “isotope effect” of the CuO$_2$ layers. This understanding receives further support from the nearly ideal structural similarity of the CuO$_2$ layers in Hg1201 and Hg1212, as opposed to variations in some of the other cuprate families (Table S3 in \cite{SM}).

\begin{figure}
	\centering{\includegraphics[clip,width=8.6cm]{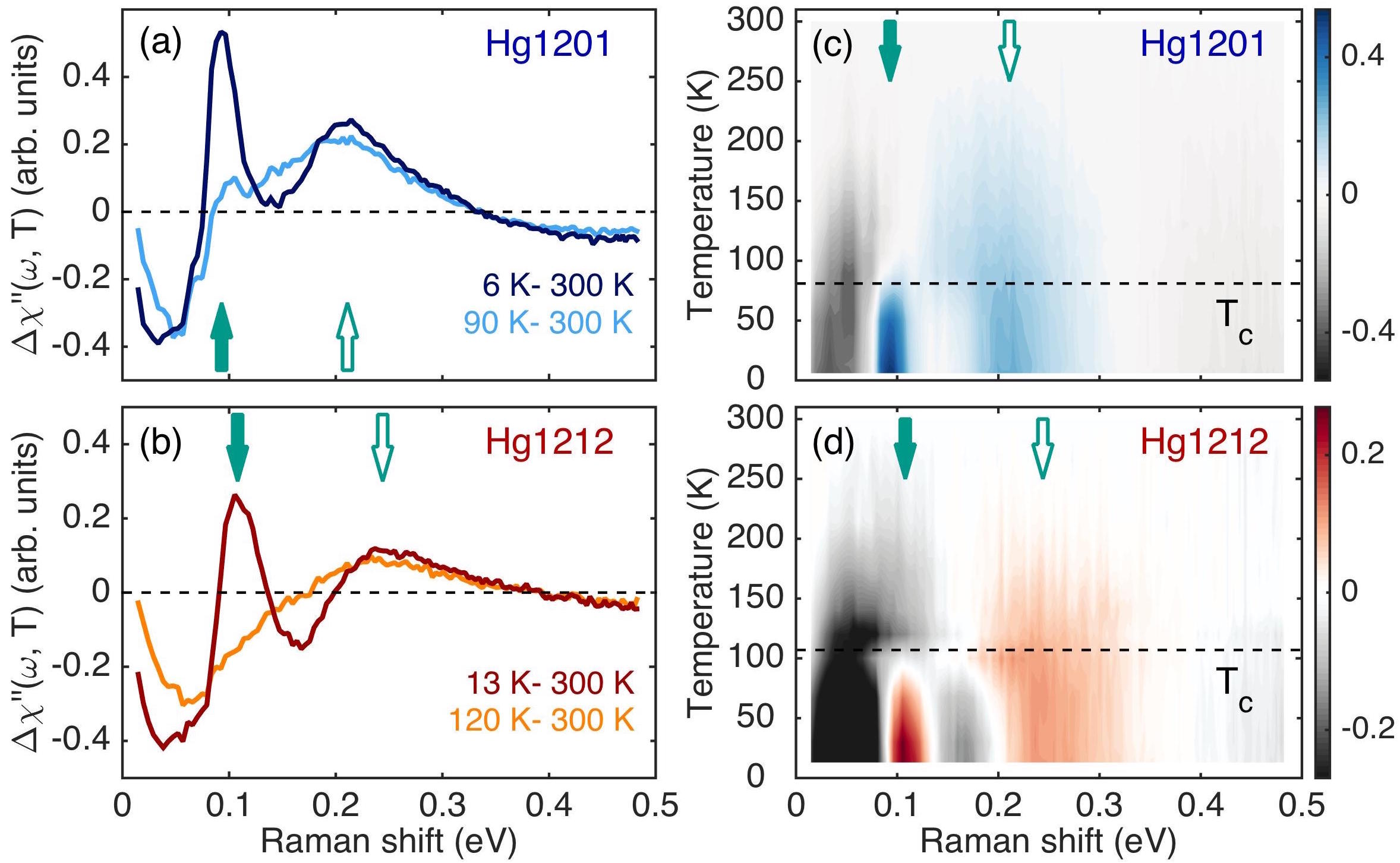}}
	\caption{$B_{1g}$ Raman spectra for Hg1201 and Hg1212. (a)-(b), Bose-factor corrected electronic Raman spectra relative to 300 K for Hg1201 and Hg1212, respectively. The $B_{1g}$ scattering geometry involves incident and scattered photons linearly polarized along the diagonals of the CuO$_2$ plaquettes, and perpendicular to each other. (c)-(d), False-color representation of Raman spectra at various $T$ relative to 300 K. Solid arrows indicate the energy of the pair-breaking peak, at 93 (108) meV for Hg1201 (Hg1212), and empty arrows the bi-paramagnon peak, at 211 (244) meV for Hg1201 (Hg1212), estimated from spectra obtained at the lowest temperature. Additional data are presented in Fig. S8 \cite{SM}.}
	\label{fig3}
\end{figure}

The pairing strength may also be measured by the superconducting gap $2\Delta_\mathrm{SC}$. To determine $2\Delta_\mathrm{SC}$ for the two systems, we have performed variable-temperature electronic Raman scattering on the same two crystals studied by RIXS. In the $B_{1g}$ scattering geometry \cite{Devereaux2007}, Raman scattering probes charge excitations from the anti-nodal regions of the Brillouin zone where the $d$-wave superconducting gap is largest. Figure 3 displays our Raman spectra taken at low temperatures after subtracting their room-temperature references, which highlights the appearance of the superconducting pair-breaking peak (at the energy of $2\Delta_\mathrm{SC}$) below $T_\mathrm{c}$. It is found that the increase in $2\Delta_\mathrm{SC}$ from Hg1201 to Hg1212 in our samples is about $16\%$, considerably smaller than the $34\%$ increase in $T_\mathrm{c}$. This difference may be partly attributed to the presumed better phase coherence \cite{Emery1995} of the superconductivity in bi-layer Hg1212, which may lead to a greater increase of $T_\mathrm{c}$ (but not necessarily of $2\Delta_\mathrm{SC}$) from Hg1201 especially in underdoped samples. The difference may also be due to the slightly different doping of our crystals, since $T_\mathrm{c}$ and $2\Delta_\mathrm{SC}$ are known to vary disproportionally with underdoping \cite{Vishik2012,Li2012,Li2013}. Near optimal doping, $2\Delta_\mathrm{SC}$ is known to be about 86 meV (Ref. \cite{Li2012}, compared to 93 meV in our sample) for Hg1201; while no measurement of $2\Delta_\mathrm{SC}$ has been reported for optimally doped Hg1212, we expect it to be somewhat smaller than 108 meV in our sample. Therefore, we estimate the increase in $2\Delta_\mathrm{SC}$ (from Hg1201 to Hg1212, at optimal doping or the same doping) to be between $16\%$ and $26\%$. The 22-30$\%$ increase in the paramagnon energies determined by RIXS is again sufficient to account for it.

In addition to the pair-breaking peak, a broad Raman peak is found to develop at a higher energy than $2\Delta_\mathrm{SC}$ upon cooling (Fig. 3, see Fig. S8 in \cite{SM} for additional data). This peak arises from excitations involving the interchange of spins (\textit{i.e.}, total $\Delta S = 0$ \cite{Jia2014}) on Cu$^{2+}$, and is known as the bi-(para)magnon peak \cite{Devereaux2007}. Several factors are known to contribute to the doping evolution of this peak \cite{Sugai2003}, including Raman resonant effects \cite{Li2013}, magnon-magnon interactions \cite{Jia2014}, doping-induced removal of spins and/or addition of itinerant carriers which disrupt the magnetic correlations \cite{Jia2014,Prelovsek1996}, etc. These complications preclude the unambiguous determination of a magnetic energy for a system by measuring the bi-paramagnon peak in doped samples only. Nevertheless, it has been reported that the bi-paramagnon energy approximately tracks the doping evolution of $2\Delta_\mathrm{SC}$ over a substantial range \cite{Li2012,Li2013,Sugai2003,Sugai2000} and, upon cooling, the peak intensity increases concurrently with the formation of Cooper pairs \cite{Li2012}. Our data in Fig. 3 reaffirm and extend these findings: the bi-paramagnon peak grows and becomes better-defined upon cooling into the superconducting state in both Hg1201 and Hg1212, and, between the two systems, the relative difference ($16\%$) in the peak energy, determined from the variation with temperature, matches precisely that of $2\Delta_\mathrm{SC}$. Combined together, these results point to the distinct possibility that the energy at which the bi-paramagnon signal changes most with temperature (especially below $T_\mathrm{c}$) is also where the associated spin excitations contribute most to the Cooper pairing. Further experimental investigation and theoretical understanding of the phenomena are warranted.

\begin{figure}
	\centering{\includegraphics[clip,width=8.6cm]{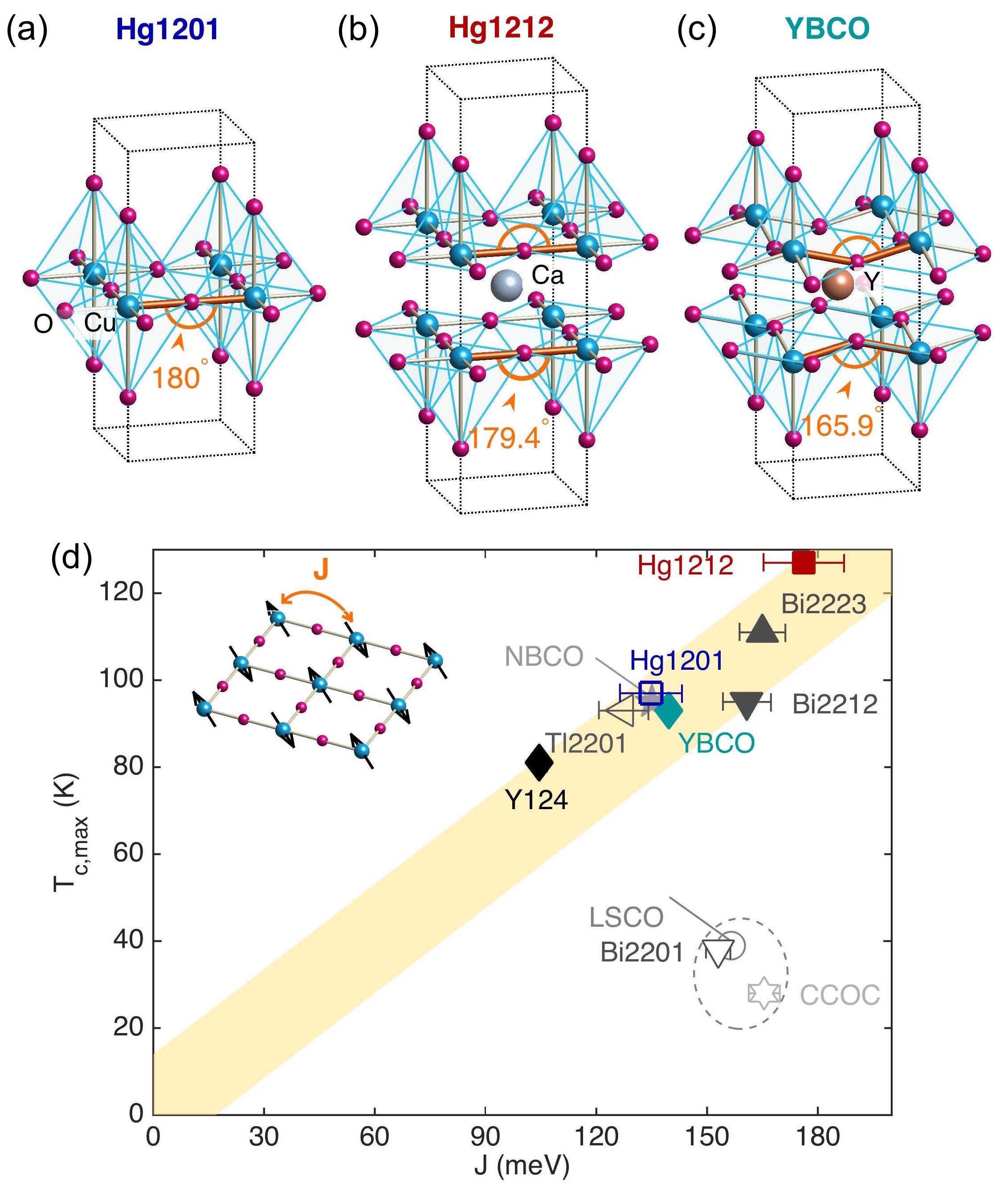}}
	\caption{(a)-(c), Schematics of CuO$_2$ layers (with apical oxygens) of Hg1201, Hg1212 and YBCO. The Cu-O-Cu bond angle (see text) is indicated in orange, and values of other compounds are summarised in Table S3 in \cite{SM}. (d), $T_\mathrm{c}$ of various cuprates at optimal or the best available doping, plotted versus $J$ extracted from neutron scattering and RIXS, see Table S4 in \cite{SM} for values and references. Thick yellow line is a guide to the eye, whereas strong outliers from this line are enclosed by a dashed ellipse. Inset illustrates $J$ between neighbouring Cu$^{2+}$ across ligand oxygen atoms. Acronyms not yet defined in the text: Bi$_2$Sr$_2$Ca$_2$Cu$_3$O$_{10+\delta}$ (Bi2223), NdBa$_2$Cu$_3$O$_{6+\delta}$ (NBCO), Tl$_2$Ba$_2$CuO$_{6+\delta}$ (Tl2201), YBa$_2$Cu$_4$O$_8$ (Y124). Error bars indicate uncertainty (1 s.d.) in the estimates of $J$.}
	\label{fig4}
\end{figure}

The significant increase in the paramagnon (and bi-paramagnon) energies from Hg1201 to Hg1212 may appear surprising at first glance. \textit{Ab initio} calculations suggest that the antiferromagnetic interaction $J$ does not vary much between the first two RP members of the Hg-family of cuprates \cite{Wan2009}. The same calculations correctly predicted the (para)magnon energies to be larger in Hg1212 than YBa$_2$Cu$_3$O$_{6+\delta}$ (YBCO) -- indeed they are observed here to be larger (Fig. 1e) than those in YBCO \cite{LeTacon2011}. This prediction is in accordance with the understanding that straighter Cu-O-Cu bonds within the CuO$_2$ layers (Fig. 4a-c) produce stronger antiferromagnetic superexchange \cite{Shimizu2003,Rocquefelte2012}, and its failure to account for the difference between Hg1201 and Hg1212 deserves a separate note. Since the on-site Coulomb repulsion $U$ in the LDA$+U$ calculations was set constant for all systems considered in Ref. \cite{Wan2009}, whether this assumption agreed with real materials may affect the accuracy of the results. We note that the charge-transfer gap was recently observed to decrease significantly from the first to the second RP member \cite{Ruan2016}, which is likely caused by their difference in the apical ions (Fig. 4a-b). Since the cuprates are charge-transfer rather than Mott insulators \cite{Lee2006}, it is reasonable to believe that the charge-transfer gap plays a similar role as $U$ and is inversely proportional to $J$ \cite{Lee2006,Ruan2016,Wang2018}. Thus, our observation can be rationalized if Hg1212 has a smaller charge-transfer gap than Hg1201 in their parent compounds. One may even ask the reversed question: Why does Bi$_2$Sr$_2$CaCu$_2$O$_{8+\delta}$ (Bi2212) not have substantially larger paramagnon energies than Bi$_2$Sr$_{2-x}$La$_x$CuO$_{6+\delta}$ (Bi2201) \cite{Dean2014,Peng2015}, even though Bi2212 has a smaller charge-transfer gap \cite{Ruan2016}? The answer, as we detail in Table S3 \cite{SM}, plausibly lies again in the bonding geometry -- Bi2212 suffers from a greater departure of the Cu-O-Cu bond angle from 180$^\circ$ than Bi2201.

The above considerations motivate us to summarize $J$ and the maximal values of $T_\mathrm{c}$ ($T_\mathrm{c,max}$, at optimal or the best available doping) for a wide spectrum of cuprates (Fig. 4d), where $J$ is determined from published inelastic neutron scattering or RIXS data (Fig. S9 in \cite{SM}). An appealing trend emerges, namely, eight different compound families approximately follow a linear relation between $T_\mathrm{c,max}$ and $J$, and a similar trend is shown in Fig. S10 \cite{SM} between $T_\mathrm{c,max}$ and the zone-boundary energy of the (para)magnons. The three outliers from this trend, Bi2201, La$_{2-x}$Sr$_x$CuO$_4$ (LSCO), and Ca$_{2-x}$Na$_x$CuO$_2$Cl$_2$ (CCOC), all have material-specific drawbacks that prevent them from reaching a higher $T_\mathrm{c,max}$, namely, the doping sites in all of them are close to the CuO$_2$ layers, which causes a large degree of disorder that is inevitably bad for high $T_\mathrm{c}$ \cite{Eisaki2004}. In addition, LSCO and CCOC have been found to lack sufficient long-range hopping integral of the itinerant carriers, which is considered instrumental for realizing a high $T_\mathrm{c}$ \cite{Pavarini2001,Peng2017}. Therefore, Fig. 4d puts forward a clearly recognizable general trend $T_\mathrm{c,max} \sim J$ across all cuprate families, although this has been partly obscured by material-specific variations and by different methods of determining $J$. Our study obviates both limitations and yields an unambiguous demonstration of the proportionality between optimally reachable $T_\mathrm{c}$ and the paramagnon energies over the entire Brillouin zone.

\begin{acknowledgments}

We are grateful to D. Betto, A. Chubukov, Ji Feng, M. Greven, G. Khaliullin, H.-H. Kim, H. Suzuki, Xiangang Wan, Fa Wang, Yayu Wang and Yuanbo Zhang for discussions. We thank Jiarui Li and Xiangpeng Luo for their help at the early stage of the project, and Prof. Shuang Jia for access to their XRD apparatus. Y.L. is grateful for financial support from the National Natural Science Foundation of China (NSFC, Grants No. 11888101 and No. 11874069) and Ministry of Science and Technology of China (MOST, Grant No. 2018YFA0305602). Y.Y.P. is grateful for financial support from the NSFC (Grant No. 11974029) and MOST (Grant No. 2019YFA0308401).

\end{acknowledgments}


\bibliography{Reference_Hg_magiso}

\end{document}